\documentclass[preprintnumbers,amsmath,amssymb,superscriptaddress]{revtex4}
\usepackage{tikz}
\usepackage{bm}
\usepackage{graphicx}
\usepackage{chemarrow}
\usepackage{multirow}

\begin{document}

\title{Induced mirror symmetry breaking via
\textbf{template}-controlled copolymerization: theoretical insights}

\author{Celia Blanco}
\email{blancodtc@cab.inta-csic.es} \affiliation{Centro de
Astrobiolog\'{\i}a (CSIC-INTA), Carretera Ajalvir Kil\'{o}metro 4,
28850 Torrej\'{o}n de Ardoz, Madrid, Spain}
\author{David Hochberg}
\email{hochbergd@cab.inta-csic.es} \affiliation{Centro de
Astrobiolog\'{\i}a (CSIC-INTA), Carretera Ajalvir Kil\'{o}metro 4,
28850 Torrej\'{o}n de Ardoz, Madrid, Spain}

\begin{abstract}
A chemical equilibrium model of template-controlled copolymerization
is presented for describing the outcome of the experimental induced
desymmetrization scenarios recently proposed by Lahav and coworkers.
\end{abstract}

\maketitle

It is an empirical fact that mirror symmetry is broken in all known
biological systems, where processes crucial for life such as
replication, imply chiral supramolecular structures, sharing the
same chiral sign (homochirality). These chiral structures are
proteins, composed of aminoacids almost exclusively found as the
left-handed enantiomers (S), also DNA, and RNA polymers and sugars
with chiral building blocks composed by right-handed (R)
monocarbohydrates.

\begin{figure}[h]
\centering
\includegraphics[width=0.48\textwidth]{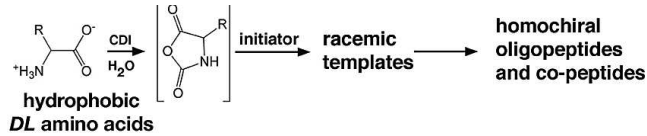}
\caption{\label{scheme} The scheme proposed in Ref.
\cite{Weissbuch2009} leading to regio-enantioselection within
racemic $\beta$-sheet templates.}
\end{figure}

One scenario for the transition from prebiotic racemic chemistry to
chiral biology suggests that homochiral peptides must have appeared
before the onset of the primeval enzymes
\cite{Bada,AGK,GAK,AG,Orgel}. However, the polymerization of racemic
mixtures (1:1 proportions) of monomers in ideal solutions typically
yields chains composed of random sequences of both the left and
right handed repeat units following a binomial distribution
\cite{Guijarro}. This statistical problem has been overcome recently
by the experimental demonstration of the generation of amphiphilic
peptides of homochiral sequence, that is, of a single chirality,
from racemic compositions. This route consists of two steps: (1) the
formation of racemic parallel or anti-parallel $\beta$-sheets either
in aqueous solution or in 3-D crystals \cite{Weissbuch2009} during
the polymerization of racemic hydrophobic $\alpha$-amino acids
followed by (2) an enantioselective controlled polymerization
reaction
\cite{Zepik2002,Nery,Nery2,Rubinstein2007,Rubinstein2008,Illos,Illos2}
(Fig. \ref{scheme}). This process leads to racemic or
mirror-symmetric mixtures of isotactic oligopeptides where the
chains are composed from amino acid residues of a single handedness.
Furthermore, when racemic mixtures of different amino acid species
were polymerized, isotactic co-peptides of homochiral sequence were
generated. Here a host or majority species $(R_0,S_0)$, together
with a given number $m$ of minority amino acid species
$(R_1,S_1),(R_2,S_2),...(R_m,S_m)$ (supplied with lesser abundance)
were employed. The guest (S) and (R) molecules are
enantioselectively incorporated into the chains of the (S) and (R)
peptides, respectively, however the former are
\textit{stochastically} distributed within the homochiral chains. As
a combined result of these two effects, the sequence of the
co-peptide S and R chains will differ from each other, resulting in
non-racemic mixtures of co-peptide polymer chains:
\textit{non-enantiomeric pairs} of chains are thus formed. By
considering the sequences of these peptide chains, a statistical
departure from the racemic composition of the library of the peptide
chains is created which varies with chain length $N$ and with the
relative concentrations of the host/guest monomers used in the
polymerization \cite{Nery,Nery2}.
The mechanism has some features in common with the scenarios
proposed by Green\cite{Green}, Eschenmoser\cite{BME} and
Siegel\cite{Siegel} in which a limited supply of material results in
a stochastic mirror symmetry breaking process.

To address the general scenario for the generation of libraries of
diastereoisomeric mixtures of peptides in accord with that proposed
in Ref.\cite{Nery}, consider a model with a host amino acid species
and $m$ guest amino acids. We assume as given the prior formation of
the initial templates or $\beta$-sheets, and are concerned
exclusively with the subsequent random polymerization reactions
(step (2)). The underlying nonlinear template control is implicit
throughout the discussion.

We consider stepwise additions and dissociations of single monomers
from one end of the (co)polymer chain, considered as a strand within
the $\beta$-sheet. It is reasonable to regard the $\beta$-sheet in
equilibrium with the free monomer pool\cite{Gonen}.

\footnotetext[18]{Reports a stochastic simulation of two concurrent
processes: 1) an irreversible condensation of activated amino acids
and 2) reversible formation of racemic $\beta$-sheets of alternating
homochiral strands, treated as a one-dimensional problem. These
architectures lead to the formation of chiral peptides whose
isotacticity increases with length.}


From detailed balance, each individual monomer attachment or
dissociation reaction is in equilibrium. This holds for closed
equilibrium systems in which the free monomers are
depleted/replenished by the templated polymerization. Then we can
compute the equilibrium concentrations of all the (co)-polymers in
terms of equilibrium constants $K_i$ and the free monomer
concentrations. The equilibrium concentration of an $S$-type
copolymer chain of length $n_0 + n_1 + n_2 + ... + n_m= N$ made up
of $n_j$ molecules $S_j$ is given by $p^S_{n_0,n_1,...,n_m} =
(K_0s_0)^{n_0}(K_1s_1)^{n_1}...(K_ms_m)^{n_m}/K_0$, where $s_j
=[S_j]$ \cite{Markvoort}. Similarly for the concentration of an
$R$-type copolymer chain of length $n'_0 + n'_1 + n'_2 + ... + n'_m=
N$ made up of $n'_j$ molecules $R_j$: $p^R_{n'_0,n'_1,...,n'_m} =
(K_0r_0)^{n'_0}(K_1r_1)^{n'_1}...(K_m r_m)^{n'_m}/K_0$, where $r_j
=[R_j]$.

The number of different $S$-type copolymers of length $l$ with $n_j$
molecules of type $S_j$ is given by the multinomial coefficient.
Hence the total concentration of the $S$-type copolymers of length
$l$ is given by
\begin{eqnarray}\label{contot}
p_l^S &=& \sum_{n_0+n_1+...+n_m=l} \left(
                                   \begin{array}{c}
                                     l \\
                                     n_0,n_1,...,n_m \\
                                   \end{array}
                                 \right)p^S_{n_0,n_1,...,n_m}
= \frac{1}{K_0}(K_0 s_0 + K_1s_1 + ... + K_ms_m)^{l},
\end{eqnarray}
which follows from the multinomial theorem \cite{nist}. We calculate
the number of each type $S_j$ of $S$-monomer present in the
$S$-copolymer of length equal to $l$, for any $0 \leq j \leq m$:
\begin{eqnarray}\label{number}
&&s_j(p^S_l) = \sum_{n_0+n_1+...+n_m=l} \left(
                                   \begin{array}{c}
                                     l \\
                                     n_0,n_1,...,n_m \\
                                   \end{array}
                                 \right)n_j p^S_{n_0,n_1,...,n_m}
= s_j \frac{\partial }{\partial s_j} p_l^S = \frac{K_j}{K_0}s_j l
(K_0 s_0 + K_1s_1 + ... + K_ms_m)^{l-1}.
\end{eqnarray}
Then we need to know the total amount of the $S$-type monomers bound
within the $S$-type copolymers, from the dimer on up to a maximum
chain length $N$. Using Eq.\ref{number} for the $jth$ type of amino
acid, this is given by
\begin{equation}\label{totamount}
s_j(p^S_{Tot}) = \sum_{l=2}^N s_j(p^S_l) \rightarrow \frac{K_j}{K_0}
s_j \frac{a(2 - a)}{(1-a)^2},
\end{equation}
the final expression holds in the limit $N \rightarrow \infty $
provided that $a = (K_0 s_0 + K_1 s_1 + ... + K_m s_m) < 1$. This
must be the case, otherwise the system would contain an infinite
number of molecules \cite{Markvoort}. Similar considerations hold
for the $R$-sector, and the total amount of $R$ monomers inside $R$
type copolymers for the $jth$ amino acid, is given by
$r_j(p^R_{Tot}) = \frac{K_j}{K_0} r_j\frac{b(2 - b)}{(1-b)^2}$ where
$b = (K_0 r_0 + K_1 r_1 + ... + K_m r_m) < 1$.
From this we obtain the mass balance equations which hold for both
enantiomers of the host and guest amino acids, and is our key
result:
\begin{equation}\label{equationsrj}
s_j + \frac{K_j}{K_0} s_j \frac{a(2 - a)}{(1-a)^2} = {s_j}_{tot},
\qquad r_j + \frac{K_j}{K_0} r_j \frac{b(2 - b)}{(1-b)^2} =
{r_j}_{tot}.
\end{equation}
These equations express the fact that each type of enantiomer is
either free, or is else bound inside a (co)polymer strand within the
template.

The problem then consists in the following: given the total
concentrations of all the $m+1$ enantiomers
$\{{s_j}_{tot},{r_j}_{tot}\}_{j=0}^m$, and the $K_i$ we calculate
the free monomer concentrations $\{s_j, r_j\}_{j=0}^m$ from solving
Eqs. (\ref{equationsrj}). Denote by $s_{0_{tot}} + ... + s_{m_{tot}}
+ r_{0_{tot}} + ... + r_{m_{tot}}= c_{tot}$ the total system
concentration. From the solutions we can calculate e.g., the
equilibrium concentrations of homochiral copolymers of any specific
sequence or composition as well as the resultant enantiomeric excess
for homochiral chains of length $l$ composed of the host (majority)
amino acid: $ee_l = \frac{(r_0)^l - (s_0)^l}{(r_0)^l + (s_0)^l}$.
When there are no guest aminoacids, i.e., for $m=0$, and when the
majority species is supplied in racemic proportions
${s_0}_{tot}:{r_0}_{tot}= 1:1$, then $ee_l$ must be zero: there will
be no mirror symmetry breaking. So we turn to the scenario of
Ref\cite{Nery} and consider the influence of a single guest species,
$m=1$ being sufficient for our purposes.

\begin{figure}[h]
\centering
\includegraphics[width=0.42\textwidth]{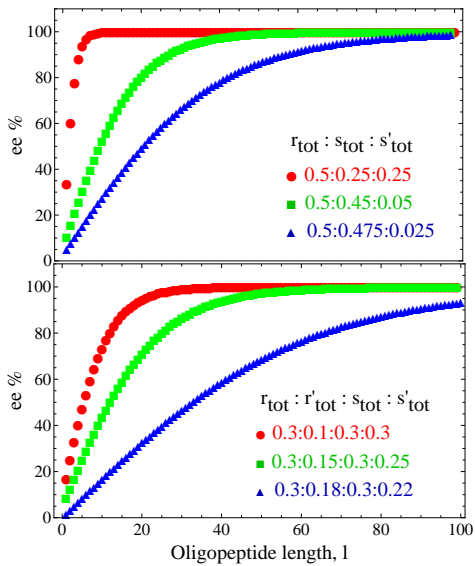}
\caption{\label{ee3comp}Calculated $ee$ values versus chain length
$l$ from solving Eqs. (\ref{equationsrj}). Top: non-racemic host
$r_{tot}>s_{tot}$ and one guest aminoacid $s'_{tot}$ $(m=1)$ and
three monomer starting compositions (in moles)
$r_{tot}:s_{tot}:s'_{tot}=0.5:0.25:0.25$ (filled circles),
$0.5:0.45:0.05$ (squares) and $0.5:0.475:0.025$ (triangles) for the
equilibrium constant $K=1M^{-1}$ and the total monomer concentration
$c_{tot}=1M$. Compare to Fig. 13 of Ref.\cite{Nery}. Bottom: racemic
host $r_{tot}=s_{tot}$ and $m=1$ guest $r'_{tot},s'_{tot}$. Starting
compositions $r_{tot}:r'_{tot}:s_{tot}:s'_{tot}=0.3:0.1:0.3:0.3$
(filled circles), $0.3:0.15:0.3:0.25$ (squares) and
$0.3:0.18:0.3:0.22$ (triangles) for four monomers.}
\end{figure}
We first use our mass balance equations to calculate $ee_l$ for the
same initial compositions of the monomers as reported in
\cite{Nery}. This is shown in top of Fig. \ref{ee3comp}. We consider
a single equilibrium constant $K_0=K_1=K=1 M^{-1}$  for sake of
simplicity, and the total system concentration, $c_{tot}=1M$. The
enantiomeric excess increases when increasing the amount of guest
species $s'_{tot}$, obtaining a maximal symmetry breaking for the
case shown with equal amounts of majority and minority S-molecules:
$s_{tot}=s'_{tot}$. In the limit as $s'_{tot}\rightarrow 0$ we tend
towards a racemic situation, so decreasing the amount of the
minority or guest species is equivalent to approaching the racemic
state, manifested through ever smaller values of $ee_l$ for fixed
$l$ (top to bottom sequence of curves). The $ee_l$ increases
monotonically with the chain length $l$ in all cases. The behavior
of the $ee_l$ demonstrates quite well the induced symmetry breaking
mechanism proposed in Ref.\cite{Nery}.

The solutions of the mass balance equations (\ref{equationsrj}) can
be used to evaluate the average chain lengths as functions of
initial monomer compositions and the equilibrium constants.  The
average chain lengths of the $S$-type copolymers $<l_S>$, composed
of random sequences of the $S_j$ type monomers, and that of the
$R$-type copolymers $<l_R>$ composed of random sequences of the
$R_j$ type monomers, are derived in the Supplementary Information.
Results for the $m=1$ three monomer cases are shown there in Table
I. There is a marked increase in the average chain length when
increasing $K$, we moreover observe how the average chain length
corresponding to each monomer species increases when increasing its
own starting proportion. In the case of additives of only one
handedness (three monomer case) and for the different compositions
considered ($r_{tot}:s_{tot}:s'_{tot}=0.5:0.25:0.25$,
$0.5:0.45:0.05$, $0.5:0.475:0.025$) the average chain length for the
$S$-type copolymers and the $R$-type polymers will be the same. This
follows since $K$ is the same for both monomer types and the amount
of $S$-type and $R$-type molecules in the starting compositions is
the same, $r_{tot}=s_{tot}+s'_{tot}$, so the average chain length
must be the same: $<l_S>=<l_R>$.

By a further example, we carry out an analysis for the case of one
guest $m=1$ and all four enantiomers, treating a majority species
$R,S$ in strictly racemic proportions and a single guest amino acid
$R',S'$ in various relative proportions. We solve Eq.
(\ref{equationsrj}) and then calculate $ee_l$ for the different
chain lengths $l$ for three different starting monomer compositions.
In Fig. \ref{ee3comp} (bottom) we show the results obtained from
calculating $ee_l$ for $K=1 M^{-1}$ and $c_{tot}=1M$. The behavior
is qualitatively similar to that previously commented, the greater
the relative disproportion of the minority species
$r'_{tot},s'_{tot}$, the greater is the enantiomeric excess. Values
for the average chain lengths are calculated for four molecules,
with the abundances
$r_{tot}:r'_{tot}:s_{tot}:s'_{tot}=0.3:0.1:0.3:0.3$ and
$r_{tot}:r'_{tot}:s_{tot}:s'_{tot}=0.3,0.14,0.3:0.26$, and are
displayed in Table II in Supplementary Information, where other
choices for the $K_i$ and $c_{tot}$ are employed (see Tables
III-VI).

In summary, we consider a multinomial sample space for the
distribution of equilibrium concentrations of homochiral copolymers
formed via template control. We deduce mass balance equations for
the enantiomers of the individual amino acid species, and their
solutions are used to evaluate the sequence-dependent copolymer
concentrations, in terms of the total species concentrations.
Measurable quantities signalling the degree of mirror symmetry
breaking such as the $ee$ and average chain lengths are evaluated.
This approach provides a quantitative basis for the
template-controlled induced desymmetrization mechanisms advocated by
Lahav and coworkers
\cite{Zepik2002,Nery,Nery2,Rubinstein2007,Rubinstein2008,Illos,Illos2}.

We are indebted to Meir Lahav for suggesting a mathematical approach
to this problem. CB has a Calvo Rod\'{e}s scholarship from INTA. DH
acknowledges a grant AYA2009-13920-C02-01 from the MICINN and forms
part of the COST Action CM0703 ``Systems Chemistry".

\footnotesize{
\bibliography{rsc} 

\begin{thebibliography}{20}
\expandafter\ifx\csname natexlab\endcsname\relax\def\natexlab#1{#1}\fi
\expandafter\ifx\csname bibnamefont\endcsname\relax
  \def\bibnamefont#1{#1}\fi
\expandafter\ifx\csname bibfnamefont\endcsname\relax
  \def\bibfnamefont#1{#1}\fi
\expandafter\ifx\csname citenamefont\endcsname\relax
  \def\citenamefont#1{#1}\fi
\expandafter\ifx\csname url\endcsname\relax
  \def\url#1{\texttt{#1}}\fi
\expandafter\ifx\csname urlprefix\endcsname\relax\def\urlprefix{URL }\fi
\providecommand{\bibinfo}[2]{#2}
\providecommand{\eprint}[2][]{\url{#2}}

\bibitem[{\citenamefont{Bada and Miller}(1987)}]{Bada}
\bibinfo{author}{\bibfnamefont{J.}~\bibnamefont{Bada}} \bibnamefont{and}
  \bibinfo{author}{\bibfnamefont{S.}~\bibnamefont{Miller}},
  \bibinfo{journal}{Biosystems} \textbf{\bibinfo{volume}{20}},
  \bibinfo{pages}{21} (\bibinfo{year}{1987}).

\bibitem[{\citenamefont{Avetisov et~al.}(1985)\citenamefont{Avetisov,
  Goldanskii, and Kuzmin}}]{AGK}
\bibinfo{author}{\bibfnamefont{V.}~\bibnamefont{Avetisov}},
  \bibinfo{author}{\bibfnamefont{V.}~\bibnamefont{Goldanskii}},
  \bibnamefont{and} \bibinfo{author}{\bibfnamefont{V.}~\bibnamefont{Kuzmin}},
  \bibinfo{journal}{Dokl. Akad. Nauk USSR} \textbf{\bibinfo{volume}{115}},
  \bibinfo{pages}{282} (\bibinfo{year}{1985}).

\bibitem[{\citenamefont{Goldanskii et~al.}(1986)\citenamefont{Goldanskii,
  Avetisov, and Kuzmin}}]{GAK}
\bibinfo{author}{\bibfnamefont{V.}~\bibnamefont{Goldanskii}},
  \bibinfo{author}{\bibfnamefont{V.}~\bibnamefont{Avetisov}}, \bibnamefont{and}
  \bibinfo{author}{\bibfnamefont{V.}~\bibnamefont{Kuzmin}},
  \bibinfo{journal}{FEBS Lett.} \textbf{\bibinfo{volume}{207}},
  \bibinfo{pages}{181} (\bibinfo{year}{1986}).

\bibitem[{\citenamefont{Avetisov and Goldanskii}(1996)}]{AG}
\bibinfo{author}{\bibfnamefont{V.}~\bibnamefont{Avetisov}} \bibnamefont{and}
  \bibinfo{author}{\bibfnamefont{V.}~\bibnamefont{Goldanskii}},
  \bibinfo{journal}{Proc. Natl. Acad. Sci. USA} \textbf{\bibinfo{volume}{93}},
  \bibinfo{pages}{11435} (\bibinfo{year}{1996}).

\bibitem[{\citenamefont{Orgel}(1992)}]{Orgel}
\bibinfo{author}{\bibfnamefont{L.}~\bibnamefont{Orgel}},
  \bibinfo{journal}{Nature} \textbf{\bibinfo{volume}{358}},
  \bibinfo{pages}{203} (\bibinfo{year}{1992}).

\bibitem[{\citenamefont{Guijarro and Yus}(2009)}]{Guijarro}
\bibinfo{author}{\bibfnamefont{A.}~\bibnamefont{Guijarro}} \bibnamefont{and}
  \bibinfo{author}{\bibfnamefont{M.}~\bibnamefont{Yus}},
  \emph{\bibinfo{title}{{T}he {O}rigin of {C}hirality in the {M}olecules of
  {L}ife}} (\bibinfo{publisher}{{RSC} {P}ublishing},
  \bibinfo{address}{{C}ambridge}, \bibinfo{year}{2009}), \bibinfo{edition}{1st}
  ed.

\bibitem[{\citenamefont{Weissbuch et~al.}(2009)\citenamefont{Weissbuch, Illos,
  Bolbach, and Lahav}}]{Weissbuch2009}
\bibinfo{author}{\bibfnamefont{I.}~\bibnamefont{Weissbuch}},
  \bibinfo{author}{\bibfnamefont{R.}~\bibnamefont{Illos}},
  \bibinfo{author}{\bibfnamefont{G.}~\bibnamefont{Bolbach}}, \bibnamefont{and}
  \bibinfo{author}{\bibfnamefont{M.}~\bibnamefont{Lahav}},
  \bibinfo{journal}{Acc. Chem. Res.} \textbf{\bibinfo{volume}{42}},
  \bibinfo{pages}{1128} (\bibinfo{year}{2009}).

\bibitem[{\citenamefont{Zepik et~al.}(2002)\citenamefont{Zepik, Shavit, Tang,
  Jensen, Kjaer, Bolbach, Leiserowitz, Weissbuch, and Lahav}}]{Zepik2002}
\bibinfo{author}{\bibfnamefont{H.}~\bibnamefont{Zepik}},
  \bibinfo{author}{\bibfnamefont{E.}~\bibnamefont{Shavit}},
  \bibinfo{author}{\bibfnamefont{M.}~\bibnamefont{Tang}},
  \bibinfo{author}{\bibfnamefont{T.}~\bibnamefont{Jensen}},
  \bibinfo{author}{\bibfnamefont{K.}~\bibnamefont{Kjaer}},
  \bibinfo{author}{\bibfnamefont{G.}~\bibnamefont{Bolbach}},
  \bibinfo{author}{\bibfnamefont{L.}~\bibnamefont{Leiserowitz}},
  \bibinfo{author}{\bibfnamefont{I.}~\bibnamefont{Weissbuch}},
  \bibnamefont{and} \bibinfo{author}{\bibfnamefont{M.}~\bibnamefont{Lahav}},
  \bibinfo{journal}{Science} \textbf{\bibinfo{volume}{295}},
  \bibinfo{pages}{1266} (\bibinfo{year}{2002}).

\bibitem[{\citenamefont{Nery et~al.}(2005)\citenamefont{Nery, Bolbach,
  Weissbuch, and Lahav}}]{Nery}
\bibinfo{author}{\bibfnamefont{J.}~\bibnamefont{Nery}},
  \bibinfo{author}{\bibfnamefont{G.}~\bibnamefont{Bolbach}},
  \bibinfo{author}{\bibfnamefont{I.}~\bibnamefont{Weissbuch}},
  \bibnamefont{and} \bibinfo{author}{\bibfnamefont{M.}~\bibnamefont{Lahav}},
  \bibinfo{journal}{Chem. Eur. J.} \textbf{\bibinfo{volume}{11}},
  \bibinfo{pages}{3039} (\bibinfo{year}{2005}).

\bibitem[{\citenamefont{Nery et~al.}(2007)\citenamefont{Nery, Eliash, Bolbach,
  Weissbuch, and Lahav}}]{Nery2}
\bibinfo{author}{\bibfnamefont{J.}~\bibnamefont{Nery}},
  \bibinfo{author}{\bibfnamefont{R.}~\bibnamefont{Eliash}},
  \bibinfo{author}{\bibfnamefont{G.}~\bibnamefont{Bolbach}},
  \bibinfo{author}{\bibfnamefont{I.}~\bibnamefont{Weissbuch}},
  \bibnamefont{and} \bibinfo{author}{\bibfnamefont{M.}~\bibnamefont{Lahav}},
  \bibinfo{journal}{Chirality} \textbf{\bibinfo{volume}{19}},
  \bibinfo{pages}{612} (\bibinfo{year}{2007}).

\bibitem[{\citenamefont{Rubinstein et~al.}(2007)\citenamefont{Rubinstein,
  Eliash, Bolbach, Weissbuch, and Lahav}}]{Rubinstein2007}
\bibinfo{author}{\bibfnamefont{I.}~\bibnamefont{Rubinstein}},
  \bibinfo{author}{\bibfnamefont{R.}~\bibnamefont{Eliash}},
  \bibinfo{author}{\bibfnamefont{G.}~\bibnamefont{Bolbach}},
  \bibinfo{author}{\bibfnamefont{I.}~\bibnamefont{Weissbuch}},
  \bibnamefont{and} \bibinfo{author}{\bibfnamefont{M.}~\bibnamefont{Lahav}},
  \bibinfo{journal}{Angew. Chem. Int. Ed.} \textbf{\bibinfo{volume}{46}},
  \bibinfo{pages}{3710} (\bibinfo{year}{2007}).

\bibitem[{\citenamefont{Rubinstein et~al.}(2008)\citenamefont{Rubinstein,
  Clodic, Bolbach, Weissbuch, and Lahav}}]{Rubinstein2008}
\bibinfo{author}{\bibfnamefont{I.}~\bibnamefont{Rubinstein}},
  \bibinfo{author}{\bibfnamefont{G.}~\bibnamefont{Clodic}},
  \bibinfo{author}{\bibfnamefont{G.}~\bibnamefont{Bolbach}},
  \bibinfo{author}{\bibfnamefont{I.}~\bibnamefont{Weissbuch}},
  \bibnamefont{and} \bibinfo{author}{\bibfnamefont{M.}~\bibnamefont{Lahav}},
  \bibinfo{journal}{Chem. Eur. J.} \textbf{\bibinfo{volume}{14}},
  \bibinfo{pages}{10999} (\bibinfo{year}{2008}).

\bibitem[{\citenamefont{Illos et~al.}(2008)\citenamefont{Illos, Bisogno,
  Clodic, Bolbach, Weissbuch, and Lahav}}]{Illos}
\bibinfo{author}{\bibfnamefont{R.}~\bibnamefont{Illos}},
  \bibinfo{author}{\bibfnamefont{F.}~\bibnamefont{Bisogno}},
  \bibinfo{author}{\bibfnamefont{G.}~\bibnamefont{Clodic}},
  \bibinfo{author}{\bibfnamefont{G.}~\bibnamefont{Bolbach}},
  \bibinfo{author}{\bibfnamefont{I.}~\bibnamefont{Weissbuch}},
  \bibnamefont{and} \bibinfo{author}{\bibfnamefont{M.}~\bibnamefont{Lahav}},
  \bibinfo{journal}{J. Am. Chem. Soc.} \textbf{\bibinfo{volume}{130}},
  \bibinfo{pages}{8651} (\bibinfo{year}{2008}).

\bibitem[{\citenamefont{Illos et~al.}(2010)\citenamefont{Illos, Clodic,
  Bolbach, Weissbuch, and Lahav}}]{Illos2}
\bibinfo{author}{\bibfnamefont{R.}~\bibnamefont{Illos}},
  \bibinfo{author}{\bibfnamefont{G.}~\bibnamefont{Clodic}},
  \bibinfo{author}{\bibfnamefont{G.}~\bibnamefont{Bolbach}},
  \bibinfo{author}{\bibfnamefont{I.}~\bibnamefont{Weissbuch}},
  \bibnamefont{and} \bibinfo{author}{\bibfnamefont{M.}~\bibnamefont{Lahav}},
  \bibinfo{journal}{Orig. Life Evol. Biosph.} \textbf{\bibinfo{volume}{40}},
  \bibinfo{pages}{51} (\bibinfo{year}{2010}).

\bibitem[{\citenamefont{Green and Garetz}(1984)}]{Green}
\bibinfo{author}{\bibfnamefont{M.~M.} \bibnamefont{Green}} \bibnamefont{and}
  \bibinfo{author}{\bibfnamefont{B.~A.} \bibnamefont{Garetz}},
  \bibinfo{journal}{Tetrahedron Letters} \textbf{\bibinfo{volume}{25}},
  \bibinfo{pages}{2831} (\bibinfo{year}{1984}).

\bibitem[{\citenamefont{Bolli et~al.}(1997)\citenamefont{Bolli, R.Micura, and
  Eschenmoser}}]{BME}
\bibinfo{author}{\bibfnamefont{M.}~\bibnamefont{Bolli}},
  \bibinfo{author}{\bibnamefont{R.Micura}}, \bibnamefont{and}
  \bibinfo{author}{\bibfnamefont{A.}~\bibnamefont{Eschenmoser}},
  \bibinfo{journal}{Chem. Biol.} \textbf{\bibinfo{volume}{4}},
  \bibinfo{pages}{309} (\bibinfo{year}{1997}).

\bibitem[{\citenamefont{Siegel}(1998)}]{Siegel}
\bibinfo{author}{\bibfnamefont{J.}~\bibnamefont{Siegel}},
  \bibinfo{journal}{Chirality} \textbf{\bibinfo{volume}{10}},
  \bibinfo{pages}{24} (\bibinfo{year}{1998}).

\bibitem[{\citenamefont{Wagner et~al.}(2011)\citenamefont{Wagner, Rubinov, and
  Ashkenasy}}]{Gonen}
\bibinfo{author}{\bibfnamefont{N.}~\bibnamefont{Wagner}},
  \bibinfo{author}{\bibfnamefont{B.}~\bibnamefont{Rubinov}}, \bibnamefont{and}
  \bibinfo{author}{\bibfnamefont{G.}~\bibnamefont{Ashkenasy}},
  \bibinfo{journal}{ChemPhysChem} \textbf{\bibinfo{volume}{12}},
  \bibinfo{pages}{2771} (\bibinfo{year}{2011}).

\bibitem[{\citenamefont{Markvoort et~al.}(2011)\citenamefont{Markvoort, ten
  Eikelder, Hilbers, de~Greef, and Meijer}}]{Markvoort}
\bibinfo{author}{\bibfnamefont{A.}~\bibnamefont{Markvoort}},
  \bibinfo{author}{\bibfnamefont{H.}~\bibnamefont{ten Eikelder}},
  \bibinfo{author}{\bibfnamefont{P.}~\bibnamefont{Hilbers}},
  \bibinfo{author}{\bibfnamefont{T.}~\bibnamefont{de~Greef}}, \bibnamefont{and}
  \bibinfo{author}{\bibfnamefont{E.}~\bibnamefont{Meijer}},
  \bibinfo{journal}{Nat. Commun.}  (\bibinfo{year}{2011}).

\bibitem[{nis(technical report)}]{nist}
\bibinfo{type}{Tech. Rep.}, \bibinfo{institution}{National Institute of
  Standards and Technology, http://dlmf.nist.gov/26} (\bibinfo{year}{technical
  report}).

\end{thebibliography}
\bibliographystyle{rsc} }

\newpage

\center{\Large{\bf{Supplementary Information}}}

\section{\label{sec:betasheet} $\beta$-sheet controlled copolymerization}

%
\begin{figure}[h]
\centering
\includegraphics[width=0.58\textwidth]{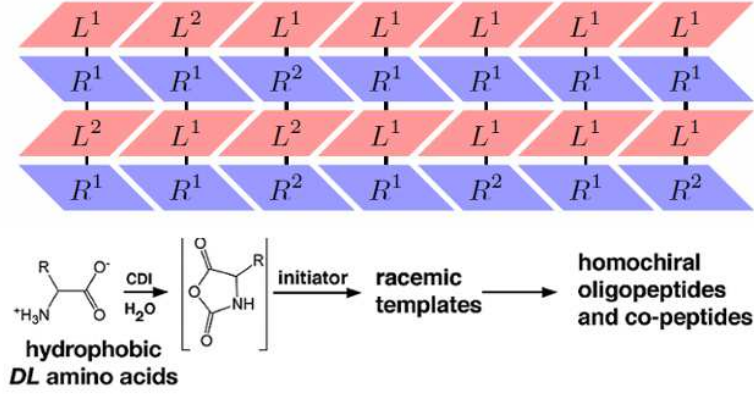}
\caption{\label{bigscheme} Regio-enantioselection within racemic
$\beta$-sheet templates.}
\end{figure}

The proposed regio-enantioselection within racemic beta sheets is
graphically illustrated by Fig \ref{bigscheme}. For sake of
simplicity, we consider a host majority species $(L_1,R_1)$ and a
minority guest species $(L_2,R_2$) of amino acids both provided in
ideally racemic proportions. The amino acids of a given handedness
attach to sites of the same chirality within the growing beta sheet
leading to the polymerization of oligomer strands of a single
chirality, in the alternating fashion as depicted. The vertical line
segments denote hydrogen bonds between adjacent strands. Since the
polymerization in any given strand is random and the guest molecules
are less abundant than the hosts, the former will attach in a random
fashion, leading to independent uncorrelated random sequences in
each strand. The overall effect leads to non-enantiomeric pairs of
chiral copolymers, so mirror symmetry is broken in a stochastic
manner.

\section{\label{sec:chainlength}Average chain lengths}

%
We can calculate the average copolymer chain lengths as functions of
initial monomer compositions ${s_j}_{tot},{r_j}_{tot}$, for the
$jth$ species, $0 \leq j \leq m$, and the equilibrium constants
$K_j$, using the solutions of our mass balance equations:
\begin{equation}\label{equationsrj}
s_j + \frac{K_j}{K_0}s_j \frac{a(2 - a)}{(1-a)^2} = {s_j}_{tot},
\qquad r_j + \frac{K_j}{K_0}r_j \frac{b(2 - b)}{(1-b)^2} =
{r_j}_{tot},
\end{equation}
where $a=K_0s_0+K_1s_1+...+K_ms_m < 1$ and
$b=K_0r_0+K_1r_1+...+K_mr_m < 1$.

The ensemble-averaged chain lengths afford an alternative measure of
the degree of mirror symmetry breaking resulting from the
desymmetrization process discussed in \cite{Nery}. There are a
number of relevant and interesting averages one can define and
calculate. The average chain lengths, starting from the dimers,  of
the $S$-type copolymers, composed of random sequences of the $S_j$
type monomers, and that of the $R$-type copolymers composed of
random sequences of the $R_j$ type monomers are given by:

\begin{eqnarray}\label{lSbarm}
<l_S> = \frac{\sum_{l=2}^N
(s_0(p^S_l)+s_1(p^S_l)+...+s_m(p^S_l))}{\sum_{l=2}^N p^S_l}
\rightarrow \frac{(s_0 + \frac{K_1}{K_0} s_1 +...+\frac{K_m}{K_0}
s_m)\frac{a(2-a)}{(1-a)^2}}{\frac{a^2}{(1-a)K_0}} =\frac{2-a}{1-a},
\end{eqnarray}
\begin{eqnarray}\label{lSbarm}
<l_R> = \frac{\sum_{l=2}^N
(r_0(p^R_l)+r_1(p^R_l)+...+r_m(p^R_l))}{\sum_{l=2}^N p^R_l}
\rightarrow \frac{(r_0 + \frac{K_1}{K_0} r_1 +...+\frac{K_m}{K_0}
r_m)\frac{b(2-b)}{(1-b)^2}}{\frac{b^2}{(1-b)K_0}} =\frac{2-b}{1-b},
\end{eqnarray}
respectively.  We also obtain an expression for the average length
of the polymer chains composed exclusively by the $S_j$ or $R_j$
monomers for a given fixed amino acid type $j$:
\begin{eqnarray}\label{ls0bar}
<l_S^{s_j}> = \frac{\sum_{l=2}^N s_j(p^{S(s_j)}_l)}{\sum_{l=2}^N
p^{S(s_j)}_l}  = \frac{\sum_{l=2}^N \frac{K_j}{K_0}s_jl(K_j
s_j)^{l-1}}{\sum_{l=2}^N \frac{(K_j s_j)^l}{K_0}} \rightarrow
\frac{\frac{(s_jK_j)^2(2-K_js_j)}{(1-K_js_j)^2}}{\frac{(K_js_j)^2}{(1-K_js_j)}}
 = \frac{2-K_js_j}{1-K_js_j},
\end{eqnarray}
\begin{eqnarray}\label{ls0bar}
<l_R^{r_j}> = \frac{\sum_{l=2}^N r_j(p^{R(r_j)}_l)}{\sum_{l=2}^N
p^{R(r_j)}_l}  = \frac{\sum_{l=2}^N \frac{K_j}{K_0}r_jl(K_j
r_j)^{l-1}}{\sum_{l=2}^N \frac{(K_j r_j)^l}{K_0}} \rightarrow
\frac{\frac{(r_jK_j)^2(2-K_jr_j)}{(1-K_jr_j)^2}}{\frac{(K_jr_j)^2}{(1-K_jr_j)}}
 = \frac{2-K_jr_j}{1-K_jr_j}.
\end{eqnarray}
To complete the list, we can calculate the chain length averaged
over all the copolymers in the system:
\begin{eqnarray}\label{lbarm}
< l > &=& \frac{\sum_{l=2}^N (s_0(p^S_l)+s_1(p^S_l)+...+s_m(p^S_l)+
r_0(p^R_l)+r_1(p^R_l)+...+r_m(p^R_l))}{\sum_{l=2}^N (p^S_l+p^R_l)}\nonumber\\
&\rightarrow& \frac{(s_0 + \frac{K_1}{K_0}s_1 +...+
\frac{K_m}{K_0}s_m)\frac{a(2-a)}{(1-a)^2}+
(r_0 + \frac{K_1}{K_0}r_1 +...+  \frac{K_m}{K_0}r_m)\frac{b(2-b)}{(1-b)^2}}{\frac{a^2}{(1-a)K_0}+\frac{b^2}{(1-b)K_0}}\nonumber\\
&=&\frac{a^2(2-a)(1-b)^2+b^2(2-b)(1-a)^2}{a^2(1-b)^2(1-a)+b^2(1-b)(1-a)^2}.
\end{eqnarray}
The right-hand most expressions $(\rightarrow)$ hold in the limit of
$N \rightarrow \infty$ and for $a<1$ and $b<1$.\\

In the following, we first consider the simplest case of $m=1$ guest
and equal equilibrium constants $K_0=K_1=K$. In the case of
additives of only one handedness (chiral additives,
$r_{tot}:s_{tot}:s'_{tot}$), and for the three different cases
considered in the Communication ($0.5:0.25:0.25$, $0.5:0.45:0.05$
and $0.5:0.475:0.025$) the average chain length for the $S$-type
copolymers and the $R$-type polymers will be the same, see Table
\ref{averagelk3comp}. This follows since the equilibrium constant is
the same for both monomer types and the amount of $S$-type and
$R$-type molecules in the starting compositions is the same
$r_{tot}=s_{tot}+s'_{tot}$, so the total average chain length must
be the same: $<l_S>=<l_R>=<l>$. In the particular case of
$r_{tot}:s_{tot}:s'_{tot}=0.5:0.25:0.25$, that is, for the same
starting amounts $s_{tot} = s'_{tot}$, the average length for the
chains exclusively composed of $S$ or $S'$ is the also same:
$<l_S^s>=<l_S^{s'}>$ (fifth and sixth columns in Table
\ref{averagelk3comp}). We can appreciate a clear increase in the
average chain length when increasing $K$ (top to bottom rows), we
observe moreover that the average chain length corresponding to each
monomer species increases when increasing its starting proportion;
see Table
\ref{averagelk3comp}, from left to right in the groups.\\

\begin{table*}[t!]
\small
  \caption{\ Average chain lengths for the three different starting compositions as a function of $K$ for $c_{tot}=1 M$}
  \label{averagelk3comp}
  \renewcommand\tabcolsep{1pt}
  \begin{tabular*}{\textwidth}{@{\extracolsep{\fill}}llllllllllllllll}
    \hline
     & \multicolumn{5}{c}{$r_{tot}:s_{tot}:s'_{tot}=0.5:0.25:0.25$} & \multicolumn{5}{c}{$r_{tot}:s_{tot}:s'_{tot}=0.5:0.45:0.05$} & \multicolumn{5}{c}{$r_{tot}:s_{tot}:s'_{tot}=0.5:0.475:0.025$} \\
    $K (M^{-1})$ & $<l>$ & $<l_S>$ & $<l_R>$ & $<l_S^s>$ & $<l_S^{s'}>$ & {$<l>$} & $<l_S>$ & $<l_R>$ & $<l_S^s>$ & $<l_S^{s'}>$ & {$<l>$} & $<l_S>$ & $<l_R>$ & $<l_S^s>$ & $<l_S^{s'}>$\\
    \hline
    $1$ & $2.37$ & $2.37$ & $2.37$ & $2.15$ & $2.15$ & $2.37$ & $2.37$ & $2.37$ & $2.32$ & $2.03$ & $2.37$ & $2.37$ & $2.37$ & $2.34$ & $2.01$\\
    $5$ & $3.16$ & $3.16$ & $3.16$ & $2.37$ & $2.37$ & $3.16$ & $3.16$ & $3.16$ & $2.93$ & $2.06$ & $3.16$ & $3.16$ & $3.16$ & $3.04$ & $2.03$\\
    $10$& $3.79$ & $3.79$ & $3.79$ & $2.47$ & $2.47$ & $3.79$ & $3.79$ & $3.79$ & $3.37$ & $2.07$ & $3.79$ & $3.79$ & $3.79$ & $3.56$ & $2.03$\\
    $50$& $6.52$ & $6.52$ & $6.52$ & $2.69$ & $2.69$ & $6.52$ & $6.52$ & $6.52$ & $4.80$ & $2.09$ & $6.52$ & $6.52$ & $6.52$ & $5.51$ & $2.04$\\
    $100$& $8.59$ & $8.59$ & $8.59$ & $2.77$ & $2.77$& $8.59$ & $8.59$ & $8.59$ & $5.57$ & $2.10$ & $8.59$ & $8.59$ & $8.59$ & $6.71$ & $2.05$\\
    $500$& $17.32$ & $17.32$ & $17.32$ & $2.88$ & $2.88$& $17.32$ & $17.32$ & $17.32$ & $7.44$ & $2.10$ & $17.32$ & $17.32$ & $17.32$ & $10.24$ & $2.05$\\
    $1000$& $23.87$ & $23.87$ & $23.87$ & $2.92$ & $2.92$& $23.87$ & $23.87$ & $23.87$ & $8.18$ & $2.11$ & $23.87$ & $23.87$ & $23.87$ & $11.92$ & $2.05$\\
         \hline
  \end{tabular*}
\end{table*}

In the particular case of
$r_{tot}:r'_{tot}:s_{tot}:s'_{tot}=0.3:0.1:0.3:0.3$, that is the
same starting amounts of $r$, $s$ and $s'$, the average chain length
for the chains exclusively composed of $s$ or $s'$ is the same,
$<l_S^s>=<l_S^{s'}>$. Numerical results for the cases
$r_{tot}:r'_{tot}:s_{tot}:s'_{tot}=0.3:0.1:0.3:0.3$ and
$r_{tot}:r'_{tot}:s_{tot}:s'_{tot}=0.3,0.14,0.3:0.26$  are shown in
Table \ref{averagelk4comp}.

We consider the effect of different equilibrium constants $K_0 \neq
K_1$ and a much smaller total system concentration $c_{tot}=10^{-3}
M$ in Table \ref{averagelk5comp}. The dependence on varying
$c_{tot}$ for fixed but distinct equilibrium constants $K_0 \neq
K_1$ is displayed in Table \ref{averagelc6comp}. These should be
compared to the previous Table \ref{averagelk3comp}, since they
refer to the same starting monomer compositions as used in that
Table. Finally Tables \ref{averagelk7comp} and \ref{averagelc8comp}
have been calculated for the same starting compositions as Table
\ref{averagelk4comp} and can be compared with the latter.

\begin{table*}[t]
\small
  \caption{\ Average chain lengths for the two different starting compositions as a function of $K$ for $c_{tot}=1 M$}
  \label{averagelk4comp}
  \renewcommand\tabcolsep{1pt}
  \begin{tabular*}{\textwidth}{@{\extracolsep{\fill}}lllllllllllllllllll}
    \hline
    & \multicolumn{7}{c}{$r_{tot}:r'_{tot}:s_{tot}:s'_{tot}=0.3:0.1:0.3:0.3$} & \multicolumn{7}{c}{$r_{tot}:r'_{tot}:s_{tot}:s'_{tot}=0.3:0.14:0.3:0.26$}\\
    $K (M^{-1})$ & {$<l>$} & $<l_S>$ & $<l_R>$ & $<l_S^s>$ & $<l_S^{s'}>$ & $<l_R^r>$ & $<l_R^{r'}>$ & {$<l>$} & $<l_S>$ & $<l_R>$ & $<l_S^s>$ & $<l_S^{s'}>$ & $<l_R^r>$ & $<l_R^{r'}>$\\
    \hline
    $1$ & $2.38$ & $2.42$ & $2.31$ & $2.17$ & $2.17$ & $2.21$ & $2.06$ & $2.37$ & $2.40$ & $2.33$ & $2.18$ & $2.15$ & $2.20$ & $2.08$ & $$\\
    $5$ & $3.18$ & $3.30$ & $3.00$ & $2.40$ & $2.40$ & $2.60$ & $2.14$ & $3.16$ & $3.25$ & $3.07$ & $2.42$ & $2.35$ & $2.54$ & $2.20$ & $$\\
    $10$& $3.82$ & $4.00$ & $3.56$ & $2.50$ & $2.50$ & $2.84$ & $2.18$ & $3.80$ & $3.92$ & $3.66$ & $2.54$ & $2.43$ & $2.74$ & $2.25$ & $$\\
    $50$& $6.57$ & $7.00$ & $6.00$ & $2.71$ & $2.71$ & $3.50$ & $2.25$ & $6.54$ & $6.82$ & $6.22$ & $2.80$ & $2.62$ & $3.23$ & $2.35$ & $$\\
    $100$& $8.64$ & $9.26$ & $7.84$ & $2.78$ & $2.78$& $3.78$ & $2.27$ & $8.61$ & $9.00$ & $8.15$ & $2.88$ & $2.68$ & $3.42$ & $2.38$ & $$\\
    $500$& $17.41$ & $18.83$ & $15.65$ & $2.89$ & $2.89$& $4.32$ & $2.30$ & $17.35$ & $18.24$ & $16.34$ & $3.02$ & $2.78$ & $3.76$ & $2.42$ & $$\\
    $1000$& $24.00$ & $26.00$ & $21.50$ & $2.92$ & $2.92$& $4.49$ & $2.31$ & $23.91$ & $25.17$ & $22.48$ & $3.06$ & $2.80$ & $3.86$ & $2.44$ & $$\\
         \hline
  \end{tabular*}
\end{table*}

\begin{table*}[t!]
\small
  \caption{\ Average chain lengths for the three different starting compositions as a function of $K_0$ for $K_1=K_0/2$ and $c_{tot}=10^{-3} M$}
  \label{averagelk5comp}
  \renewcommand\tabcolsep{1pt}
  \begin{tabular*}{\textwidth}{@{\extracolsep{\fill}}llllllllllllllll}
    \hline
     & \multicolumn{5}{c}{$r_{tot}:s_{tot}:s'_{tot}=0.5:0.25:0.25$} & \multicolumn{5}{c}{$r_{tot}:s_{tot}:s'_{tot}=0.5:0.45:0.05$} & \multicolumn{5}{c}{$r_{tot}:s_{tot}:s'_{tot}=0.5:0.475:0.025$} \\
    $K_0 (M^{-1})$ & $<l>$ & $<l_S>$ & $<l_R>$ & $<l_S^s>$ & $<l_S^{s'}>$ & {$<l>$} & $<l_S>$ & $<l_R>$ & $<l_S^s>$ & $<l_S^{s'}>$ & {$<l>$} & $<l_S>$ & $<l_R>$ & $<l_S^s>$ & $<l_S^{s'}>$\\
    \hline
    $1$ & $2.00$ & $2.00$ & $2.00$ & $2.00$ & $2.00$ & $2.00$ & $2.00$ & $2.00$ & $2.00$ & $2.00$ & $2.00$ & $2.00$ & $2.00$ & $2.00$ & $2.00$\\
    $10$ & $2.00$ & $2.00$ & $2.00$ & $2.00$ & $2.00$ & $2.00$ & $2.00$ & $2.00$ & $2.00$ & $2.00$ & $2.00$ & $2.00$ & $2.00$ & $2.00$ & $2.00$\\
    $100$& $2.04$ & $2.04$ & $2.05$ & $2.02$ & $2.01$ & $2.05$ & $2.05$ & $2.05$ & $2.04$ & $2.00$ & $2.05$ & $2.05$ & $2.05$ & $2.04$ & $2.00$\\
    $1000$& $2.34$ & $2.31$ & $2.37$ & $2.17$ & $2.10$ & $2.36$ & $2.36$ & $2.37$ & $2.32$ & $2.02$ & $2.36$ & $2.36$ & $2.37$ & $2.34$ & $2.00$\\
    $10000$& $3.76$ & $3.73$ & $3.79$ & $2.51$ & $2.42$& $3.79$ & $3.78$ & $3.80$ & $3.40$ & $2.06$ & $3.79$ & $3.79$ & $3.79$ & $3.58$ & $2.03$\\
    $100000$& $8.57$ & $8.56$ & $8.59$ & $2.78$ & $2.75$& $8.59$ & $8.58$ & $8.59$ & $5.60$ & $2.10$ & $8.59$ & $8.59$ & $8.59$ & $6.73$ & $2.04$\\
         \hline
  \end{tabular*}
\end{table*}

\begin{table*}[t!]
\small
  \caption{\ Average chain lengths for the three different starting compositions as a function of $c_{tot}$ for $K_0=100000$ and $K_1=K_0/2$}
  \label{averagelc6comp}
  \renewcommand\tabcolsep{1pt}
  \begin{tabular*}{\textwidth}{@{\extracolsep{\fill}}llllllllllllllll}
    \hline
     & \multicolumn{5}{c}{$r_{tot}:s_{tot}:s'_{tot}=0.5:0.25:0.25$} & \multicolumn{5}{c}{$r_{tot}:s_{tot}:s'_{tot}=0.5:0.45:0.05$} & \multicolumn{5}{c}{$r_{tot}:s_{tot}:s'_{tot}=0.5:0.475:0.025$} \\
    $c_{tot} (M)$ & $<l>$ & $<l_S>$ & $<l_R>$ & $<l_S^s>$ & $<l_S^{s'}>$ & $<l>$ & $<l_S>$ & $<l_R>$ & $<l_S^s>$ & $<l_S^{s'}>$ & $<l>$ & $<l_S>$ & $<l_R>$ & $<l_S^s>$ & $<l_S^{s'}>$\\
    \hline
    $10^{-5}$& $2.34$  & $2.31$ & $2.37$ & $2.17$ & $2.10$ &      $2.36$  & $2.36$ & $2.36$ &  $2.32$ & $2.02$&   $2.36$    & $2.36$ & $2.37$ & $2.34$ & $2.01$\\
    $10^{-4}$& $3.76$  & $3.73$ & $3.79$ & $2.51$ & $2.42$ &      $3.79$ & $3.78$ &  $3.79$ & $3.40$ & $2.06$ &   $3.79$    & $3.79$ & $3.79$ & $3.58$ & $2.03$\\
    $10^{-3}$& $8.57$  & $8.56$ & $8.59$ & $2.78$ & $2.75$&       $8.59$  & $8.58$ & $8.59$ & $5.60$ & $2.09$&    $8.59$    & $8.59$ & $8.59$ & $6.73$ & $2.04$\\
    $10^{-2}$ & $23.86$  & $23.86$ & $23.87$ & $2.92$ & $2.91$&   $23.87$& $23.86$ & $23.87$ & $8.18$ & $2.11$&   $23.87$ & $23.87$ & $23.87$ & $11.93$ & $2.05$\\
    $10^{-1}$ & $72.21$  & $72.21$ & $72.21$ & $2.97$ & $2.97$&   $72.22$ & $72.22$ & $72.21$ & $9.88$ & $2.11$&   $72.24$ & $72.27$ & $72.21$ & $16.79$ & $2.05$\\
         \hline
  \end{tabular*}
\end{table*}

\begin{table*}[t]
\small
  \caption{\ Average chain lengths for the two different starting compositions as a function of $K_0$ for $K_1=K_0/2$ and $c_{tot}=10^{-3} M$}
  \label{averagelk7comp}
  \renewcommand\tabcolsep{1pt}
  \begin{tabular*}{\textwidth}{@{\extracolsep{\fill}}lllllllllllllllllll}
    \hline
    & \multicolumn{7}{c}{$r_{tot}:r'_{tot}:s_{tot}:s'_{tot}=0.3:0.1:0.3:0.3$} & \multicolumn{7}{c}{$r_{tot}:r'_{tot}:s_{tot}:s'_{tot}=0.3:0.14:0.3:0.26$}\\
    $K_0 (M^{-1})$ & {$<l>$} & $<l_S>$ & $<l_R>$ & $<l_S^s>$ & $<l_S^{s'}>$ & $<l_R^r>$ & $<l_R^{r'}>$ & {$<l>$} & $<l_S>$ & $<l_R>$ & $<l_S^s>$ & $<l_S^{s'}>$ & $<l_R^r>$ & $<l_R^{r'}>$\\
    \hline
    $1$ & $2.00$ & $2.00$ & $2.00$ & $2.00$ & $2.00$ & $2.00$ & $2.00$ &    $2.00$ & $2.00$ & $2.00$ & $2.00$ & $2.00$ & $2.00$ & $2.00$\\
    $10$ & $2.00$ & $2.00$ & $2.00$ & $2.00$ & $2.00$ & $2.00$ & $2.00$ &    $2.00$ & $2.00$ & $2.00$ & $2.00$ & $2.00$ & $2.00$ & $2.00$\\
    $100$& $2.04$ & $2.04$ & $2.03$ & $2.03$ & $2.01$ & $2.03$ & $2.00$ &    $2.04$ & $2.04$ & $2.04$ & $2.03$ & $2.01$ & $2.03$ & $2.00$\\
    $1000$& $2.33$ & $2.36$ & $2.28$ & $2.19$ & $2.12$ & $2.22$ & $2.04$ &    $2.33$ & $2.35$ & $2.30$ & $2.20$ & $2.10$ & $2.22$ & $2.05$\\
    $10000$& $3.77$ & $3.94$ & $3.53$ & $2.53$ & $2.45$& $2.88$ & $2.16$ &    $3.75$ & $3.86$ & $3.61$ & $2.58$ & $2.40$ & $2.78$ & $2.22$\\
    $100000$& $8.62$ & $9.23$ & $7.83$ & $2.80$ & $2.77$& $3.81$ & $2.27$ &    $8.58$ & $8.79$ & $8.13$ & $2.89$ & $2.67$ & $3.44$ & $2.37$\\
         \hline
  \end{tabular*}
\end{table*}

\begin{table*}[t]
\small
  \caption{\ Average chain lengths for the two different starting compositions as a function of $c_{tot}$ for $K_0=100000$ and $K_1=K_0/2$}
  \label{averagelc8comp}
  \renewcommand\tabcolsep{1pt}
  \begin{tabular*}{\textwidth}{@{\extracolsep{\fill}}lllllllllllllllllll}
    \hline
    & \multicolumn{7}{c}{$r_{tot}:r'_{tot}:s_{tot}:s'_{tot}=0.3:0.1:0.3:0.3$} & \multicolumn{7}{c}{$r_{tot}:r'_{tot}:s_{tot}:s'_{tot}=0.3:0.14:0.3:0.26$}\\
    $c_{tot} (M)$ & {$<l>$} & $<l_S>$ & $<l_R>$ & $<l_S^s>$ & $<l_S^{s'}>$ & $<l_R^r>$ & $<l_R^{r'}>$ & {$<l>$} & $<l_S>$ & $<l_R>$ & $<l_S^s>$ & $<l_S^{s'}>$ & $<l_R^r>$ & $<l_R^{r'}>$\\
    \hline
    $10^{-5}$ & $2.33$ & $2.36$ & $2.28$ & $2.19$ & $2.12$ & $2.22$ & $2.04$ &    $2.33$ & $2.35$ & $2.30$ & $2.20$ & $2.10$ & $2.22$ & $2.05$\\
    $10^{-4}$ & $3.77$ & $3.94$ & $3.53$ & $2.53$ & $2.45$ & $2.88$ & $2.16$ &    $3.75$ & $3.86$ & $3.61$ & $2.58$ & $2.40$ & $2.78$ & $2.22$\\
    $10^{-3}$& $8.62$ & $9.23$ & $7.83$ & $2.79$ & $2.77$ & $3.81$ & $2.26$ &    $8.58$ & $8.97$ & $8.13$ & $2.89$ & $2.67$ & $3.44$ & $2.37$\\
    $10^{-2}$& $23.99$ & $25.99$ & $21.50$ & $2.92$ & $2.92$ & $4.49$ & $2.31$ &    $23.90$ & $25.16$ & $22.48$ & $3.06$ & $2.80$ & $3.86$ & $2.43$\\
    $10^{-1}$& $72.58$ & $78.96$ & $64.75$ & $2.97$ & $2.97$& $4.82$ & $2.33$ &    $72.34$ & $76.33$ & $67.83$ & $3.12$ & $2.85$ & $4.05$ & $2.46$\\
         \hline
  \end{tabular*}
\end{table*}

\end{document}